# Surface Plasmon Assisted Gentle Ablation of Nanostructures by Femtosecond Oscillator

Liping Shi,[1,2,*] Bianca Iwan,[1,4] Quentin Ripault,[3] José R. C. Andrade,[1,2] Seunghwoi Han,[5] Hyunwoong Kim,[5] Willem Boutu,[3] Dominik Franz,[3] Rana Nicolas,[3] Torsten Heidenblut,[6] Carsten Reinhardt,[7] Bert Bastiaens,[8] Tamas Nagy,[1,9] Ihar Babuskin,[1,10] Uwe Morgner,[1,2,6] Seung-Woo Kim,[5] Günter Steinmeyer [10], Hamed Merdji,[3] Milutin Kovačev [1,2,†]

[1] *Institut für Quantenoptik, Leibniz Universität Hannover, Welfengarten 1, 30167 Hannover, Germany*
[2] *QUEST, Centre for Quantum Engineering and Space-Time Research, 30167 Hannover, Germany*
[3] *LIDYL, CEA, CNRS, Université Paris-Saclay, CEA Saclay 91191 Gif-sur-Yvette, France*
[4] *PULSE Center, Stanford University, California, USA*
[5] *Department of Mechanical Engineering, Korea Advanced Institute of Science and Technology (KAIST), Science Town, Daejeon, 305-701, South Korea*
[6] *Institut für Werkstoffkunde, Leibniz Universität Hannover, An der Universität 2, 30823 Garbsen, Hannover, Germany*
[7] *Laser Zentrum Hannover e.V., Hollerithallee 8, D-30419 Hannover, Germany*
[8] *Laser Physics and Nonlinear Optics, Mesa+ Institute for Nanotechnology, University of Twente, Enschede, The Netherlands*
[9] *Laser-Laboratorium Gottingen e.V., Hans-Adolf-Krebs-Weg 1, D-37077, Göttingen, Germany*
[10] *Max-Born-Institut, Max-Born-Str. 2a, D-12489 Berlin, Germany*
[*]*Corresponding author: shi@iqo.uni-hannover.de*
[†]*Corresponding author: kovacev@iqo.uni-hannover.de*

We experimentally demonstrate the use of subwavelength optical nanoantennae to assist the gentle ablation of nanostructures directly using ultralow fluence from a Ti: sapphire oscillator through the excitation of surface plasmon waves. We show that this ablation mechanism is the same for metal and dielectric. The analytical solutions of ablation threshold are in excellent agreement with the experiment estimations. Surface plasmon assisted locally enhanced ablation at nanoscale provides a method for nanomachining, manipulation and modification the nanostructures without collateral thermal damage to the materials. It is also shown that this ablation can deposit low-density high quality thin nano film.

Laser ablation is defined as the removal of material from a solid target by direct absorption of photon energy, which is of considerable interests in scientific research and industrial applications, such as laser micro-machining, deposition of thin film, and creation of new materials [1, 2]. The high ablation threshold of solids typically requires the utilization of high laser fluence operating at low repetition rate. High fluence however leads to various unwanted results especially due to the thermal effects, e. g., the collateral thermal damage around the laser spot, the formation of droplets and nanoparticulates on the deposited thin film, etc [3-5]. By operating the femtosecond laser at the intensity near the ablation threshold, it has been shown that the electrostatic ablation of solid bulk will be dominant, which eliminates these major problems caused by thermal processes [6, 7]. However, the machining nanoscale targets of interest by femtosecond pulses still suffers the serious thermal problems due to the larger surface-to-volume ratio and smaller heat capacity of the nanoscale materials. For instance, the fluence threshold for the gentle ablation of gold bulk is ~0.6 J/cm$^2$ [8], while the thermal damage threshold for a gold nanoparticle with the diameter of 100 nm is less than 0.01 J/cm$^2$ [9]. Therefore, the thermal melting and subsequent vaporization is widely believed as the dominant ablation mechanism of nanostructures by femtosecond laser.

In order to prevent the thermal issues during the manipulation of nanostructures, one approach is to employ high peak intensity while keeping the incident laser fluence at a low level. For this objective, here we demonstrate that the plasmonic nanoantennae [10] hold a promising potential for the thermal-free ablation and modification for both of metallic and dielectric nanostructures. The collective oscillation of free electrons with the skin layer of nanoantennae can produce a high field concentration in an extremely small area, which can result in several orders of magnitude enhancement of the local field strength in the vicinity of a sharp metallic tip or in the feed gap of a dimeric nanoantenna [11]. This locally enhanced near-field has been extensively employed, such as near-field scanning optical microscopy [12], photo-detectors [13], photovoltaic cells [14], thermal emitters [15], saturable absorbers [16], extreme-ultraviolet emission [17], single-molecule fluorescence enhancement [18], harmonic generation enhancement [19, 20], and optical tweezers [21]. In this letter, we demonstrate the use of nanoantennae to assist the gentle ablation and modification of dielectric and metal nanostructures with high repetition rate femtosecond oscillator, which also provides a method to fabricate thin nanofilm with high surface quality.

We first investigated the plasmonic antennae enhanced ablation of dielectric fused silica ($SiO_2$). We manufactured a gold (Au) bowtie nanoantenna by focused ion beam (FIB) milling of a gold film deposited on a sapphire substrate. A 3-nm-thick chromium (Cr) film is deposited first on the sapphire ($Al_2O_3$) acting as an adhesion layer for the gold film. The bowtie consists of tip-to-tip oriented triangles with a length of 200 nm, a height of 135 nm, and an opening angle of 40°. After fabrication the Au antenna, we coated a 5-nm $SiO_2$ nanofilm to cover the whole Au structures. A Ti: Sapphire oscillator with a repetition rate of 100 MHz, pulse duration of < 8 fs, and central wavelength of 800 nm is utilized to shine the nanostructure. The incident peak laser intensity is measured to be $2.5 \times 10^{11}$ W/cm$^2$. The laser polarization is aligned to be parallel with the longitudinal axis of the nanoantenna. For the employed laser intensity here, we can neglect the adiabatic metallization, optical breakdown and the nonlinear optical response of the dielectric nanofilm.

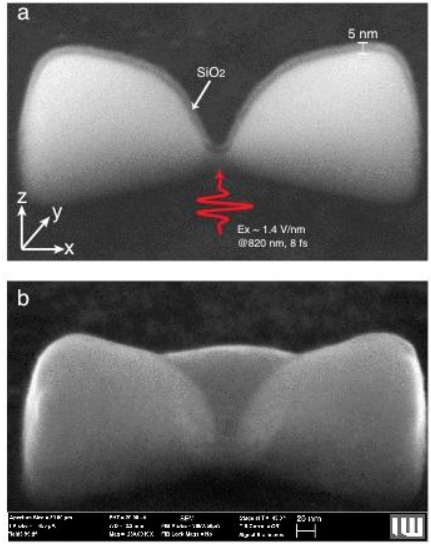

Figure 1 Schematic illustration of the experiment. The SEM images of a gold bowtie nanoantenna in perspective-view (tilted angle 45°) without (a), and with (b) 30 minutes of laser irradiation. The gold nanostructures are covered by SiO2 dielectric film.

Figure 1(a) and (b) show the scanning electron microscopy (SEM) pictures of a nanoantenna before and after $t_i$ = 30 minutes of laser irradiation. It clearly shows that in the gap between the adjacent tips, an ellipsoidal nanostructure is produced by laser illumination. To gain further insight into this laser produced nanostructure, we analyze the top-view image taken with the SEM, i.e. in x-y-plane, as shown in Fig. 2 (a). Moreover, Fig. 2 (b) depicts its SEM topography from a cross-sectional view, which is obtained by milling a cross section of this nanostructure in x-z plane with a Ga$^+$-based FIB. One can see from both views that the deposited nanostructure is not a hollow bubble. Instead, the whole gap region is filled by the deposition of material.

In order to discriminate the chemical elements of this laser-induced deposition, we apply the energy dispersive spectroscopy (EDS). The EDS element maps in x-y plane (left row) and x-z plane (right row) are shown in Fig. 2 (c-j). The dashed curves depict the outline of the nanoantennae. The analysis reveals that the chemical elements of the generated bubble-shaped nanostructure in the gap region are silicon (Fig. 2 c, d) and oxygen (Fig. 2 e, f). As aforementioned, these two elements originate from the $SiO_2$ coating, which is initially covered on the surface of the Au nanoparticles. The element distribution of Au (Fig. 2 g, h) is still excellently bowtie-shaped after laser illumination. The absence of the Au in the deposition proves that there is no ablation of the Au nanostructure itself. The distribution of Cr (Fig. 2 i, j) clearly shows the 3-nm chromium adhesion layer.

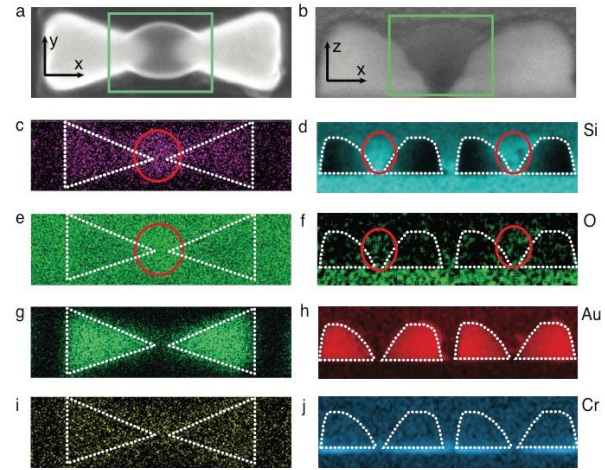

Figure 2 SEM images and EDS maps. SEM images of bowtie nanoantennas after 30 minutes of laser irradiation in (a) x-y plane and (b) in x-z plane. (c-j) EDS element maps of the laser exposed nanostructure in x-y plane (left row) and in x-z plane (right row), which shows that the generated nanostructure in the gap is SiO2, as shown inside the red circles in Fig. c, d, e and f. Note that the EDS maps in x-y plane and x-z plane are in different scale.

Our experiments conducted at few-cycle pulse duration and ultralow laser fluence allow us to exclude the thermal ablation, which is typically interpreted as the dominant ablation mechanism for long pulses [8]. The plasmonic current in the skin layer of the Au antennae and subsequent thermal diffusion can indeed heat the system to a high temperature. However, the $SiO_2$ has a higher melting point compared with Au. The absence of Au deformation confirms that its temperature is below the melting point of our hybrid system. By using the two-temperature model, we confirm that the system temperature is indeed much lower than the melting point of $SiO_2$. Moreover, the ablation here is achievable for laser fluence much far below the laser-induced explosive boiling.

The observed material removal around the nanostructure tips implies that the ablation is closely related to the resonant-plasmon-induced near-field enhancement. As shown in Fig. 3, the maximum field enhancement factor at the surface layer of $SiO_2$ tips reaches 24, corresponding to a peak intensity of $1.4\times10^{14}$ W/cm$^2$. Theoretical predictions have pointed out that the electrostatic acceleration dominates the ablation mechanism of practically any target material for sub-picosecond laser pulses at an intensity above $10^{13}$–$10^{14}$ W/cm$^2$ [6, 7]. The dielectric materials exposed in such a strong electric field will experience dramatic ionization and plasma formation at the beginning of femtosecond pulses, which is followed by plasma heating via inverse bremsstrahlung and vacuum absorption processes. In case a free electron is accelerated to a kinetic energy exceeding a sum of its escaping energy and ion binding energy, this energetic electron can pull its parent ion out of the surface. This process is called electrostatic ablation (gentle ablation), since its ablation rate is much lower than the thermal vaporization. The analytical solution of electrostatic ablation threshold is given by [6, 7]

$$I_{th} = \frac{3}{8}(\varepsilon_b + J_i)\frac{\lambda}{t_p}\frac{n_a}{2\pi}. \quad (1)$$

At laser wavelength $\lambda = 820$ nm and pulse duration $t_p = 8$ fs, by using the following parameters for $SiO_2$, $\varepsilon_b = 3.7$ eV and $J_i = 1.36$ eV meaning the atom binding energy and electron ionization potential, $n_a = 7\times10^{22}$ cm$^{-3}$ the atoms number density, the ablation threshold is calculated to be $I_{th} = 1.14\times10^{14}$ W/cm$^2$. This analytical solution is in excellent agreement with the numerical simulation of near-field enhancement, as shown in Fig. 3.

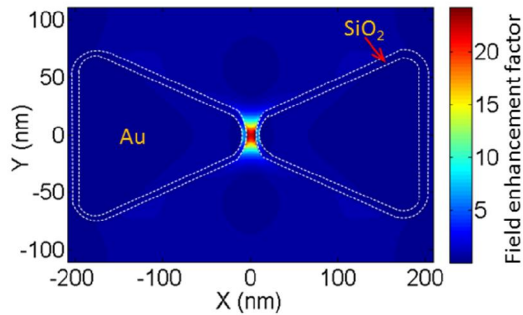

Figure 3. Numerical simulation of near-field distribution of an Au bowtie nanoantenna with 5-nm $SiO_2$ coating.

The total atom number of $SiO_2$ initially coated at both Au tips is estimated to be $N = V_0 \times n_a = 5.6\times10^6$, with $V_0 = 8\times10^4$ nm$^3$ denotes the volume of coating film at the bowtie curvature tips. As a result, the gentle ablation rate of $SiO_2$ is estimated to be $\eta = N / t_i = 3\times10^3$ s$^{-1}$, which is several orders of magnitude lower than that of thermal ablation.

The thermal ablation induced deposition of thin film has the major defects such as formation of droplets and nanoparticles. However, the relatively low ablation rate as shown in our experiments can significantly eliminate these thermal caused problems. From the SEM images in Fig. 1b we can see that the ejected $SiO_2$ accumulate in the gap region and form into a very smooth thin film. By measuring the volume of this deposited thin film $V_1 = 7.8 \times 10^5$ nm$^3$, its density is estimated to be $n_a' = 7.2 \times 10^{21}$ cm$^{-3}$, which is almost one order of magnitude less than the initial state of $SiO_2$ coating. The electrostatic ablation occurs in the time scale of subpicosecond, which is evidently shorter than the electron–phonon coupling time (2–5 ps) [23]. Therefore, the lattice remains at room temperature during the ablation time, which suggests that the electrostatic ablation is a cold process. This cold ablation might be another reason for the deposition of smooth thin film.

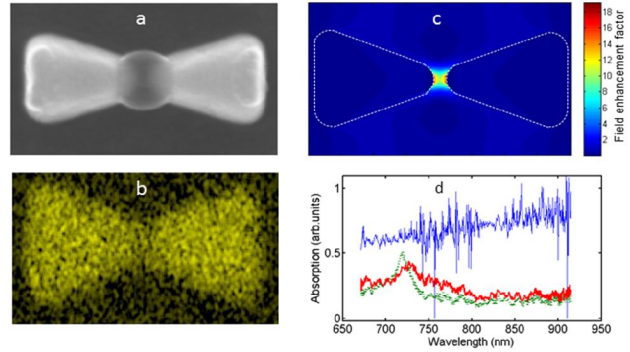

Figure 4. SEM image (a) and EDS map (b) of an Au antenna after laser irradiation. (c) Numerical simulation of near-field distribution of the prepared antenna. (d) Absorption spectra of an Au film with thickness of 50 nm (solid blue curve), plasmon resonant Au nanostructure (green curve) and the Au nanostructure after laser irradiation with the deposition of low-density thin film experiments (red curve).

For nanosecond pulses induced thermal ablation, the behavior of transparent dielectric materials is totally different from their metallic counterparts. In contrast, for the femtosecond pulses caused gentle ablation at high peak intensity, due to the ionization occurring at the surface of dielectrics, they proceed in a same way with respect to metals. In order to further verify the gentle ablation at the nanoscale materials, we also investigate the ablation of the Au nanoantennae themselves. Figure 4a, b show the SEM image and EDS map of a pure Au bowtie after 30 minutes of laser illumination by using the same laser parameters, from which we can see the same results compared to $SiO_2$, i.e. the ablation is confined at the sharpest tips and without any thermal damage to the shanks of antenna. This result further confirms that the ablation is not originated from the thermal process, because the thermal deformation is expected to appear at the nanoantennae shank, where the heat source density is maximum [24]. The parameters for Au is given by $\varepsilon_b$ = 8.2 eV, $n_a$ = 5.9×10$^{22}$ cm$^{-3}$, $\varepsilon_{esc}$ =4.8 eV the electronic work function, which is instead of $J_i$ for metals. Therefore, according to Eq. (1), the gentle ablation threshold for Au is calculated to be $I_{th}$ = 7.5×10$^{13}$ W/cm$^2$. The FDTD-based numerical simulation of this nanoantenna shows that the near-field enhancement at the Au surface layer is 18 (Fig. 4c), corresponding to the peak intensity of 8.1×10$^{13}$ W/cm$^2$, which also agrees well with the calculated ablation threshold as aforementioned. From the SEM pictures as depicted in Fig. 4a, b we can see that the gentle ablation of Au atoms also produces a smooth and low-density thin film in the gap region of dimeric antennae.

We perform the absorption spectroscopy to study the optical property of the laser-ablation-induced thin film. As a reference, we first measured the absorption spectrum of a 50-nm-thick Au film prepared by thermal evaporation (Fig. 4d, blue curve), which shows that the absorption increasing for wavelength longer than 650 nm (infrared). This is due to the free-electron behavior of Au lattice atoms [25]. For locally surface plasmon resonant nanoantennae array, the absorption peaks at the resonant wavelength at 725 nm (green curve). After laser ablation, these nanoantennae were totally threaded by the deposited Au thin film. If this deposited thin film has the same dielectric constant with respect to the initially prepared Au film, the plasmon resonance should disappear, and the absorption will increase in the infrared region. However, our measurement shows that (Fig. 4d red curve) the deposited thin film does not obviously vary the surface plasmon resonance, although these initially separated nanoantennae were connected by Au atoms. This implies that the free electron density of this laser-produced thin film is quite low, which also agrees with the indication from the SEM image, i.e. the deposited thin film is a kind of low-density Au foam.

In summary, we have demonstrated that the resonant plasmon can assist the local gentle ablation of both metallic and dielectric nanostructures directly by a femtosecond oscillator operating at ultralow laser fluence by at high repetition rate. This locally near-field enhanced cold ablation provides an effective scheme for femtosecond nanomachining and deposition of high-quality thin film. Further investigations of the optical and electrical properties of the deposited thin film with absorption spectroscopy and x-ray diffraction is likely to yield new insights into the material structure.

We would like to thank the funding support from the Deutsche Forschungsgemeinschaft (DFG) under grant number KO 3798/4-1, from the Centre for Quantum Engineering and Space-Time Research (QUEST), from the National Research Foundation of the Republic of Korea (NRF-2012R1A3A1050386), from the ANR under the grant IPEX (2014), from the LABEX PALM (ANR-10-LABX-0039) under the grants Plasmon-X (2014) and HILAC (2015). Bianca Iwan is grateful for financial support from the Swedish Research Council (637-2013-439). Liping Shi is grateful for Professor Jeremy Baumberg from University of Cambridge for stimulating discussion.